\def\papertitle{Virtual Analog Modeling of Distortion Circuits Using Neural Ordinary Differential Equations}
\def\paperauthorA{Jan Wilczek}
\def\paperauthorB{Alec Wright}
\def\paperauthorC{Emanu{\"e}l A. P. Habets}
\def\paperauthorD{Vesa V{\"a}lim{\"a}ki}
\documentclass[twoside,a4paper]{article}
\usepackage{etoolbox}
\usepackage{dafx_22a}
\usepackage{amsmath,amssymb,amsfonts,amsthm}
\usepackage{euscript}
\usepackage[T1]{fontenc}
\usepackage[utf8]{inputenc}
\usepackage{nimbusserif}
\usepackage{ifpdf}
\usepackage[english]{babel}
\usepackage{caption}
\usepackage{color}

\input glyphtounicode
\ninept

\newcounter{numauth}
\newcounter{listcnt}
\newcommand\authcnt[1]{\ifdefined#1 \stepcounter{numauth} \fi}

\newcommand\addauth[1]{
\ifdefined#1 
\stepcounter{listcnt}
\ifnum \value{listcnt}<\value{numauth}
\appto\authorslist{, #1}
\else
\appto\authorslist{~and~#1}
\fi
\fi}
\authcnt{\paperauthorB}
\authcnt{\paperauthorC}
\authcnt{\paperauthorD}
\authcnt{\paperauthorE}
\authcnt{\paperauthorF}
\authcnt{\paperauthorG}
\authcnt{\paperauthorH}
\authcnt{\paperauthorI}
\authcnt{\paperauthorJ}
\def\authorslist{\paperauthorA}
\addauth{\paperauthorB}
\addauth{\paperauthorC}
\addauth{\paperauthorD}
\addauth{\paperauthorE}
\addauth{\paperauthorF}
\addauth{\paperauthorG}
\addauth{\paperauthorH}
\addauth{\paperauthorI}
\addauth{\paperauthorJ}
\usepackage{times}
\newif\ifpdf
\ifx\pdfoutput\relax
\else
   \ifcase\pdfoutput
      \pdffalse
   \else
      \pdftrue
\fi

\ifpdf % compiling with pdflatex
  \usepackage[pdftex,
    pdftitle={\papertitle},
    pdfauthor={\authorslist},
    pdfsubject={Proceedings of the 25th International Conference on Digital Audio Effects (DAFx20in22)},
    colorlinks=false, % links are activated as color boxes instead of color text
    bookmarksnumbered, % use section numbers with bookmarks
    pdfstartview=XYZ % start with zoom=100% instead of full screen; especially useful if working with a big screen :-)
  ]{hyperref}
  \usepackage[pdftex]{graphicx}
\else % compiling with latex
  \usepackage[dvips]{epsfig,graphicx}
  \usepackage[dvips,
    pdftitle={\papertitle},
    pdfauthor={\authorslist},
    pdfsubject={Proceedings of the 25th International Conference on Digital Audio Effects (DAFx20in22)},
    colorlinks=false, % no color links
    bookmarksnumbered, % use section numbers with bookmarks
    pdfstartview=XYZ % start with zoom=100% instead of full screen
  ]{hyperref}
\fi
\usepackage[hypcap=true]{caption}
\title{\papertitle}

\threeaffiliations{
\paperauthorA \,\sthanks{This work was supported by a fellowship within the IFI programme of the German Academic Exchange Service (DAAD).}}
{\href{https://www.thewolfsound.com}{WolfSound} \\ Katowice, Poland\\
{\tt \href{mailto:jan.wilczek@thewolfsound.com}{jan.wilczek@thewolfsound.com}}
}
{\paperauthorB \ and \paperauthorD\,\sthanks{
This research is part of the activities of the Nordic Sound and Music Computing Network---NordicSMC (NordForsk project no.~86892).}}
{\href{https://www.aalto.fi/en/aalto-acoustics-lab}{Acoustics Lab}\\Dept. Signal Processing and Acoustics
 \\ Aalto University, Espoo, Finland
 \\ {\tt \href{mailto:alec.wright@aalto.fi}{alec.wright@aalto.fi}}
}
{\paperauthorC}
{\href{https://www.audiolabs-erlangen.de/}{International Audio Laboratories Erlangen}\,\sthanks{A joint institution of the Friedrich-Alexander-Universit{\"a}t Erlangen-N{\"u}rnberg (FAU) and Fraunhofer Institute for Integrated Circuits (IIS).\vspace{3mm}} \\ Erlangen, Germany \\ {\tt \href{mailto:emanuel.habets@audiolabs-erlangen.de}{emanuel.habets\\@audiolabs-erlangen.de}}
}
\usepackage{pgfplots}
\usepackage{circuitikz}
\ctikzset{bipoles/length=0.9cm}
\usepackage[printonlyused]{acronym}
\usepackage{pgf-umlsd}
\usetikzlibrary{patterns,shapes.arrows,calc,dsp,chains}
\usepgfplotslibrary{groupplots}
\usepackage{subcaption}
\usepackage{siunitx}
\usepackage{booktabs}                        % for nicer looking tables
\usepackage{makecell}
\usepackage{multirow}
\hypersetup{hidelinks}

\usepackage[sort]{cite}
\usepackage{flushend}

\newcommand{\Figure}[1]{Figure~\ref{#1}}
\newcommand{\Figures}[2]{Figures~\ref{#1}~and~\ref{#2}}
\newcommand{\Equation}[1]{Eq.~\eqref{#1}}

\newcommand{\Table}[1]{Table~\ref{#1}}

\newcommand{\Section}[1]{Section~\ref{#1}}

\newcommand{\Vector}[1]{\ensuremath{\pmb{#1}}}
\begin{document}
% more pdf-tex settings:
\ifpdf % used graphic file format for pdflatex
  \DeclareGraphicsExtensions{.png,.jpg,.pdf}
\else  % used graphic file format for latex
  \DeclareGraphicsExtensions{.eps}
\fi

%\makeatletter
%\pdfbookmark[0]{\@pdftitle}{title}
%\makeatother
\acrodef{DC}{direct current}
\acrodef{AC}{alternating current}
\acrodef{ANN}{artificial neural network}
\acrodef{BDF}{backward difference formula}
\acrodef{BPTT}{backpropagation through time}
\acrodef{DAW}{digital audio workstation}
\acrodef{DOPRI}{Dormand-Prince}
\acrodef{ESR}{error-to-signal ratio}
\acrodef{GPU}{graphics processing unit}
\acrodef{IC}{initial condition}
\acrodef{ITU}{International Telecommunication Union}
\acrodef{IVP}{initial value problem}
\acrodef{LFO}{low-frequency oscillator}
\acrodef{LSD}[LSD\textsubscript{rms}]{root mean square log spectral distance}
\acrodef{LSTM}{Long Short-Term Memory}
\acrodef{LTI}{linear time-invariant}
\acrodef{LTV}{linear time-variant}
\acrodef{MLP}{multilayer perceptron}
\acrodef{MSE}{mean square error}
\acrodef{ODE}{ordinary differential equation}
\acrodef{ODG}{Objective Difference Grade}
\acrodef{PEAQ}{Objective Measurements of Perceived Audio Quality}
\acrodef{ReLU}{Rectified Linear Unit}
\acrodef{ResBlock}{residual block}
\acrodef{ResNet}{residual network}
\acrodef{RINN}{Residual Integration Network}
\acrodef{RK}{Runge-Kutta}
\acrodef{RNN}{recurrent neural network}
\acrodef{SDG}{Subjective Difference Grade}
\acrodef{SDR}{signal-to-distortion ratio}
\acrodef{segSNR}{segmental signal-to-noise ratio}
% \acrodef{fw-segSNR}{frequency-weighted segmental signal-to-noise ratio}
\acrodef{SELU}{scaled exponential linear unit}
\acrodef{STFT}{short-time Fourier transform}
\acrodef{STN}{State Trajectory Network}
\acrodef{VA}{Virtual Analog}

\maketitle

\begin{abstract}
    Recent research in deep learning has shown that neural networks can learn differential equations governing dynamical systems. In this paper, we adapt this concept to \ac{VA} modeling to learn the \acp{ODE} governing the first-order and the second-order diode clipper. The proposed models achieve performance comparable to state-of-the-art \acp{RNN} albeit using fewer parameters. We show that this approach does not require oversampling and allows to increase the sampling rate after the training has completed, which results in increased accuracy. Using a sophisticated numerical solver allows to increase the accuracy at the cost of slower processing. \acp{ODE} learned this way do not require closed forms but are still physically interpretable.
\end{abstract}

\acresetall

\sloppy

\section{Introduction}
\label{sec:intro}

\Ac{VA} models are digital emulations of audio systems that were originally built using analog electronic or electromechanical components \cite{Zoelzer2011}. They arose out of the demand for the reproduction of characteristic tonal distortions of analog devices but with the digital stability and ease of integration with existing software. Devices modeled range from filters \cite{Parker2019} through time-varying effects \cite{Eichas2014,Kiiski2016,Wright2020}, amplifiers \cite{Wright2019a, Wright2019, Wrightetal2020}, and mechanical reverb units \cite{Parker2013} to tape or vinyl distortions \cite{Pakarinenetal2011}.

One can distinguish three general approaches to \ac{VA} modeling \cite{Kiiski2016,Wright2020}. In \emph{black-box} modeling, only the input-output relation of a system is examined and a signal model is constructed to mimic that behavior. Neural networks have successfully been applied to this kind of modeling for guitar amplifiers \cite{Wright2019a,Wright2019, Wrightetal2020}. An architecture provided in \cite{Wright2019} was chosen as a baseline for this work.

In \emph{white-box} modeling, the internal structure of the system under study is examined and used to construct an algorithm reproducing the behavior of the device. Sometimes this approach is referred to as \emph{physical modeling}. Typical methods in this category are numerical solutions of \acp{ODE} derived from electronic circuit analysis \cite{Eichas2014,Yeh2007,Yeh2008} or wave-digital filters \cite{PASPWEB2010}. 

In white-box modeling, it is assumed that analog electronic circuits are dynamical systems that can be described by differential equations. % TODO: Add citation here
However, the derivation of these equations is a difficult task that requires expert knowledge. %TODO: Citation?
Additionally, solving these equations poses a challenge because of the inherent aliasing \cite{Yeh2007}. Typical approaches to this problem, such as oversampling or local iterations of implicit solvers, increase the processing times, often preventing real-time applications.

\emph{Grey-box} modeling falls in between the two already mentioned approaches. In this methodology, we use some knowledge about the inner workings of the device under study to design a model and, subsequently, take advantage of the available signal data to adjust the model's parameters. This approach has been successfully applied to distortion circuits modeling \cite{Parker2019}, adjusting component values of white-box circuit models \cite{Esquedaetal2021}, and time-varying effects modeling \cite{Kiiski2016,Wright2020} also in conjunction with neural networks.
    
In related work, Parker et al. used a \ac{MLP} to learn the residual of a state-space system \cite{Parker2019} in a grey-box fashion. This \ac{ResNet} \cite{He2015} approach was called \ac{STN}. The authors successfully applied \ac{STN} to model analog first-order and second-order clipper circuits and an analog filter. \ac{STN} can be used at sampling rates different than the sampling rate of the training set by properly scaling the residual but the authors did not provide any results concerning this feature.

Recent research relating deep learning and \acp{ODE} has indicated that \acp{ODE} governing analog electronic circuits could be learned from data. The concept of teaching a neural network the derivative of an unknown function and then supplying the learned derivative to a numerical solver, termed ODENet, was introduced in \cite{Chen2018}. Karlsson and Svanström applied it to dynamical systems modeling \cite{Karlsson2019}. Since this approach combines data-driven neural network training (an approach typically classified as black-box) and the assumption that the modeled system is governed by an \ac{ODE} (usually associated with white-box models), it could be classified as grey-box.

The potential benefits of coupling a neural network with a numerical solver in \ac{VA} modeling are manifold. First, we could obtain an empirical \ac{ODE} that would replace complicated analytical expressions based on simplifying assumptions and often unknown, imprecise, or condition-dependent physical quantities \cite{Esquedaetal2021}. Taking this even further, we could model audio effects previously not described by \acp{ODE} using solely signal data. Second, we could possibly alleviate various problems inherent to numerical solutions of closed-form \acp{ODE} such as aliasing. Third, since an \ac{ODE} is physically interpretable, we could alter the sampling rate of the model after training, which is typically not possible with purely black-box models. Finally, by obtaining simplified, alias-free \acp{ODE}, we could add real-time performance capabilities to white-box models, thus, letting musicians and producers use previously unavailable software emulations of analog hardware.

To validate the above possible benefits, we adapted ODENet to \ac{VA} modeling. In this paper, we applied it to the first- and the second-order diode clipper. Through these examples, \emph{we confirmed all of the above potential benefits} leaving possible real-time implementation for future work. This paper discusses the ODENet implementation in the context of \ac{VA}, presents the results of experiments, and analyzes strengths and weaknesses of this approach.

This paper is organized as follows. In \Section{sec:node}, the general concept of learning a derivative is described. In \Section{sec:node_va}, the proposed extension to this approach, which facilitates \ac{VA} modeling, is discussed. In \Section{sec:modeling}, we apply the derivative-learning approach to the first-order and the second-order diode clipper circuits, compare it to the baseline, and discuss the results. Finally, \Section{sec:conclusions} recapitulates the most important conclusions and forms the perspective for future work. Audio examples of the presented models can be found under \url{https://www.thewolfsound.com/publications/dafx2022/}. The full code without the datasets can be found under \url{https://github.com/JanWilczek/va-phaser-with-neural-odes}.

\section{Neural Ordinary Differential Equations}
\label{sec:node}

An \acf{ODE} is an equation of the form
\begin{equation}
    \frac{d\Vector{y}}{dt} = f(t, \Vector{y}),
    \label{eq:ode}
\end{equation}
where $\Vector{y} = \Vector{y}(t)$ is the vector-valued \emph{unknown function} of an \emph{independent variable} $t$, here representing time, and $f(t,\Vector{y})$ is a known function of $t$ and $\Vector{y}$, which equals to the first derivative of $\Vector{y}$ over time \cite{Gockenbach2011}. An entry of the unknown function vector may, for example, represent the voltage across a capacitor inside a circuit. We refer to a vector-valued unknown function for generality and because every system of \acp{ODE} can be formulated as a system of first-order \acp{ODE} \cite{Karlsson2019}.

\begin{figure}
    \centering
    \scalebox{0.9}{\begin{tikzpicture}
    \node[dspnodeopen,dsp/label=above]              (input) {$\pmb{t}, \pmb{y}_0, f(t, \pmb{y})$};
    \node[dspfilter,right=of input, minimum width=2.5cm,xshift=1cm]                                (odesolve) {\texttt{ODESolve}};
    \node[dspnodeopen,dsp/label=above,right=of odesolve,xshift=1cm]              (output) {$[\pmb{y}_0, \pmb{y}_1, \dots, \pmb{y}_{N-1}]$};

    \draw[dspconn] (input) -- (odesolve);
    \draw[dspconn] (odesolve) -- (output);
\end{tikzpicture}}
    \caption{A general pipeline for solving \acp{ODE}. In ODENet, derivative function $f$ is replaced by a neural network.}
    \label{fig:ode_solving}
\end{figure}
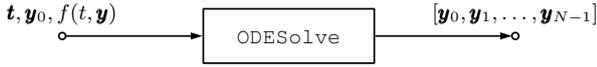

% TODO: Fix discrepancy between vector-valued y and 1D y.

In \Figure{fig:ode_solving}, one can see a generic approach to solving a system of \acp{ODE} with a numerical solver. A vector of time points $\Vector{t}$, is supplied with the derivative function $f(t, \Vector{y})$, and the \emph{initial value} $\Vector{y}_0$ (value of $\Vector{y}$ at $t = t[0]$) to a numerical solver (marked as \texttt{ODESolve}). The result is a series of estimates of the value of $\Vector{y}$ at the time points specified in $\Vector{t}$. The initial value plays a crucial role in solving most \acp{ODE} because it determines the initial state and may influence the overall dynamics of the system. 

% \begin{figure}
%     \centering
%     \input{figures/oscillator_trajectories}
%     \caption{Numerical solution of the damped harmonic oscillator equation. (top) Displacement over time for zero initial displacement and nonzero initial velocity. (bottom) Displacement over time of the same oscillator but with zero initial velocity; the system does not oscillate at all.}
%     \label{fig:oscillator}
% \end{figure}

% A vivid example can be seen in \Figure{fig:oscillator} where the oscillations of a damped harmonic oscillator for two different initial conditions are shown. A system with zero initial displacement and a nonzero initial velocity (top) results in time-varying output. A system with initial displacement and velocity equal to zero (bottom) results in a constant, zero output.

Chen et al. \cite{Chen2018} showed that a neural network used in conjunction with a numerical solver can learn $f(t,\Vector{y})$ from data. In that framework, termed ODENet, the derivative function $f$ from \Equation{eq:ode} and \Figure{fig:ode_solving} is replaced with a \emph{derivative network} $f$ which is trained based on the loss calculated using the solver's output.

Karlsson and Svanström \cite{Karlsson2019} demonstrated that a neural network can learn $f(t,\Vector{y})$ describing a simple dynamical system such as an oscillator. That indicated that a neural network can learn the derivative function of \acp{ODE} describing analog electronic circuits for \ac{VA} purposes, which could provide the benefits mentioned in Section \ref{sec:intro}.

\section{Extension for Virtual Analog Modeling}
\label{sec:node_va}

\begin{figure*}[h!]
  \center
  \input{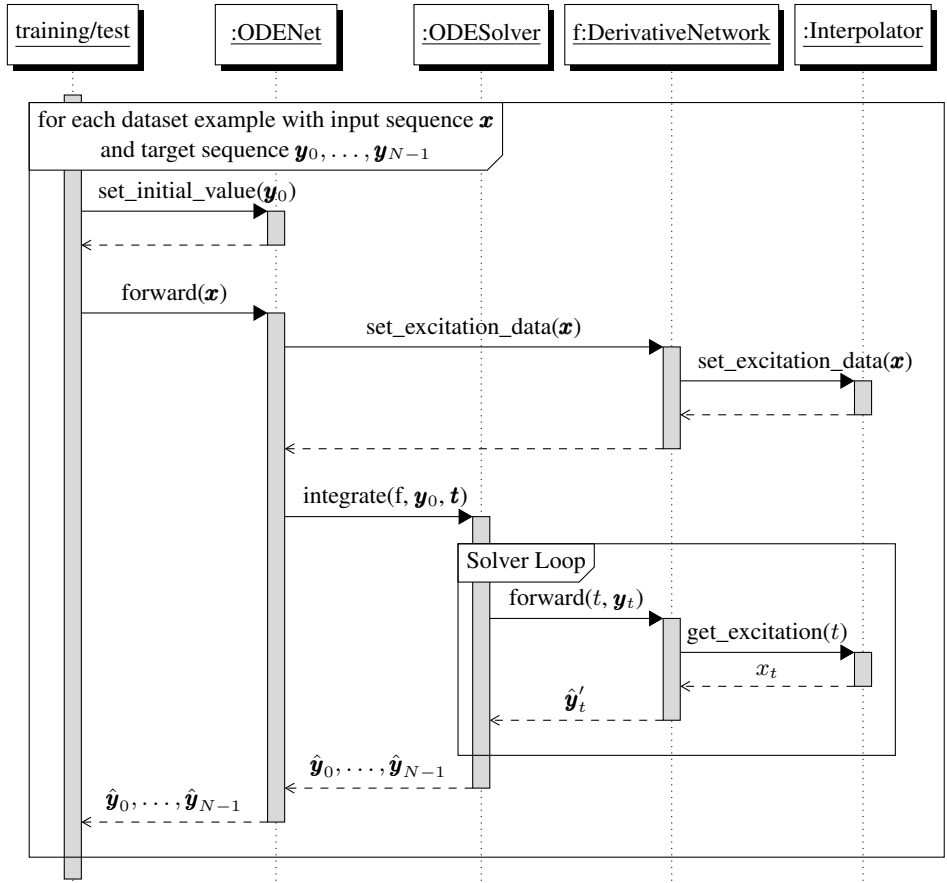}
  \caption{Sequence diagram containing the details of ODENet processing. The label ``forward'' denotes instance calls with specified arguments.}
  \label{fig:odenet_sequence_diagram}
\end{figure*}

To use ODENet for \ac{VA} modeling, a few issues must be considered: what initial value to supply, how to incorporate the input signal (e.g., a guitar signal), how to parametrize the neural network, and which solver to use.

\subsection{Overview}
\label{subsec:odenet_va_overview}

In \Figure{fig:odenet_sequence_diagram}, our implementation of the ODENet framework for \ac{VA} modeling is presented. For clarity, we omit the idea of minibatches and assume that a single example is processed at a time. One dataset example is a sequence of samples of the input signal to the modeled audio effect, $\Vector{x} = [x_0, \dots, x_{N-1}]^\text{T}$, and the corresponding target sequence (desired effect output), $\Vector{y}_0, \dots, \Vector{y}_{N-1}$. The output of ODENet is an estimated sequence, $\Vector{\hat{y}}_0, \dots, \Vector{\hat{y}}_{N-1}$, which is used to compute the loss.

\subsection{Initial Value}
\label{subsec:initial_value}

% This was already shown on the example of a damped harmonic oscillator in \Figure{fig:oscillator}. 

Supplying an initial value is marked as "set\_initial\_value($\Vector{y}_0)$" in \Figure{fig:odenet_sequence_diagram}. A proper initial value to the numerical solver plays a crucial role in obtaining a correct result of the \ac{ODE} \cite{Gockenbach2011}. An inaccurate initial condition can lead the solver to a completely incorrect part of the solution space. That is why we decided to supply the ground truth initial value for each subsequence during training, an approach known in deep learning as teacher forcing \cite{Goodfellow-et-al-2016,Peussaetal2021}. This is especially important because we split our training set into sequences of \SI{22050}{} samples. A group of sequences constitutes a minibatch. Each minibatch is processed in subsequences of \SI{2048}{} samples (after each subsequence, there is a gradient step). Thus, each subsequence is most likely to start in the middle of a waveform, so a zero value would be incorrect most of the time.

However, at test time, the framework must not be able to access ground truth information. Additionally, in contrast to the training phase, the test sequence is processed as one long sequence without minibatches. Therefore, we decided to use an all-zero vector as the initial value of the test sequence, i.e., $\Vector{y}_0 = \Vector{0}$.

\subsection{Excitation}
\label{subsec:excitation}

% The input signal to an audio effect unit could be, for example, a guitar signal. The modification applied to this signal by the audio effect is what determines its character. For the purpose of this article, we will assume that the input signal is one-dimensional, i.e., consists of a single channel of sampled data.

In \acp{ODE} modeled with ODENet prior to this work, only a simple, fixed-cosine excitation had been considered \cite{Karlsson2019}. In the context of \ac{VA} modeling, the excitation is the input signal to the modeled audio effect. We provide the input signal directly to the neural network. The derivative network uses linear interpolation to obtain the input signal value at time points specified by the solver. In other words, when the solver calls the network $f$ as $f(t,\Vector{y}_t)$ for some $t$ and $\Vector{y}_t$, the network linearly interpolates the discrete input signal $\pmb{x}$, to obtain an estimate of $x(t)$, $x_t$, and then uses $x_t$ and $\Vector{y}_t$ as its input to output the estimate of the derivative $\hat{\Vector{y}}'_t$. This can be seen in the "Solver Loop" in \Figure{fig:odenet_sequence_diagram}.

\subsection{Neural Network Parametrization}
\label{subsec:nn_parametrization}

Although the ODENet framework as a whole may be perceived as a form of a \ac{RNN} (because it may use previous outputs to compute subsequent outputs), the derivative itself only maps the time and the state vector to the derivative value as shown in \Equation{eq:ode}. This property narrows down the space of possible parametrizations of the derivative network to feedforward networks, by our choice, \acp{MLP}. 

% We denote \ac{MLP} sizes by $l_1 \times l_2 \times \dots \times l_N$, where $l_i$ is the number of units in the $i$-th layer with $l_1$ referring to the input layer and $l_N$ referring to the output layer. All layers are fully connected and apply the same nonlinearity to their output apart from the last one, which is a simple weighting layer. Additionally, $l_1 = l_N + 1$ to account for the excitation signal incorporated at the input. This can be seen 

\begin{figure}[t]
    \centering
    \scalebox{0.42}{\input{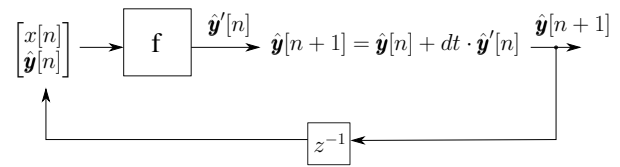}}
    \caption{ODENet processing using the forward Euler scheme.}
    \label{fig:odenet_fe_diagram}
\end{figure}

In \Figure{fig:odenet_fe_diagram}, we show how ODENet works under the forward Euler integration scheme \cite{Yeh2007}. The current input sample and current output sample are provided to the derivative network that estimates the current derivative value. This derivative value is used to estimate the next output sample, which is retained for the subsequent iteration. 

The difference between the processing in \Figure{fig:odenet_fe_diagram} and \ac{STN} (\ac{ResNet}) processing is that the latter would use a ``current input, previous output'' neural network input, not the ``previous input, previous output'' (see Figure 2 in \cite{Parker2019}). Generally speaking, while \ac{STN} has a fixed numerical scheme, ODENet allows for a flexible choice of a numerical solver and a use of different solvers at training and at test time. Additionally, ODENet learns the continuous derivative, whereas the \ac{STN} learns its discrete approximation. Finally, ODENet and \ac{STN} have significantly different learning dynamics. From our observations, the training of \ac{STN} requires severe regularization in the form of learning rate schedules \cite{Smith2018}.

\subsection{Numerical Solvers}

Applying numerical solvers to audio processing is challenging. In time-stepping methods, one needs to multiply the derivative by the time step size or its fraction at each step \cite{Gockenbach2011}. In audio, the step size is very small, resulting in large derivative variations, numerical errors, and slow neural network learning. To combat these issues, we scaled the underlying \ac{ODE} by the sampling rate so that the time step size is 1. We adequately scaled down the state vector passed to the derivative network to not alter the derivative learned.
 
Another challenge related to numerical solvers is their real-time performance. Explicit solvers can become unstable when applied to certain kinds of \acp{ODE} with an insufficient step size (i.e., sample rate). For example, Yeh et al. needed to use 38x oversampling with the forward Euler scheme when solving the analytical \ac{ODE} of the first-order diode clipper \cite{Yeh2007}. Implicit solvers, which iterate locally to reduce the error and, thus, provide very accurate solutions, have unpredictable running times unless one bounds the number of iterations per step. They also need oversampling, albeit smaller \cite{Yeh2007}. Explicit versus implicit choice is an example of a broader design issue of choosing a numerical solver: typically, with increased accuracy comes increased processing time. In this work, we show that oversampling is not needed when using a derivative learned from data, but the solver choice is still relevant for the model's accuracy.

\section{Modeling Distortion Circuits}
\label{sec:modeling}

This section presents the analog systems that were modeled, the modeling procedure, and the obtained results. First, the training procedure is outlined. Second, the first-order diode clipper is presented and the results of modeling it discussed with distinction between model quality and accuracy at different sampling rates. Finally, the second-order diode clipper is treated in a similar way.

\subsection{Training}
\label{subsec:training}

% \subsubsection{Loss Function}

After \cite{Wright2020,Wright2019a,Wright2019,Wrightetal2020}, this work used the following combined loss function for neural network training of the first-order diode clipper
\begin{equation}
  \mathcal{E}(y, \hat{y}) = \mathcal{E}_\text{ESR}(y_\text{p}, \hat{y}_\text{p}) + \mathcal{E}_\text{DC}(y, \hat{y}),
  \label{eq:final_loss_function}
\end{equation}
where $y$ is the target signal, $\hat{y}$ is the estimated signal, ``p'' in the subscript marks signals that were pre-emphasized with a first-order high-pass filter of the form \cite{Wright2020,Wright2019a}
\begin{equation}
  H(z) = 1 - 0.85 z^{-1},
  \label{eq:preemphasis_filter}
\end{equation}
\ac{ESR}, $\mathcal{E}_\text{ESR}$, is given by
\begin{equation}
  \mathcal{E}_\text{ESR}(y_\text{p}, \hat{y}_\text{p}) = \frac{\sum_{n=0}^{N-1} (y_\text{p}[n] - \hat{y}_\text{p}[n])^2}{\sum_{n=0}^{N-1} (y_\text{p}[n])^2},
  \label{eq:esr}
\end{equation}
and the \ac{DC} loss term, $\mathcal{E}_\text{DC}$, is given by
\begin{equation}
  \mathcal{E}_\text{DC}(y, \hat{y}) = \frac{\left(\frac{1}{N} \sum_{n=0}^{N-1} (y[n] - \hat{y}[n])\right)^2}{\frac{1}{N} \sum_{n=0}^{N-1} (y[n])^2}.
\end{equation}
For the second-order diode clipper, we used plain $\mathcal{E}_\text{ESR}$ given by \Equation{eq:esr} not to suppress the \ac{DC} component in the second state of the circuit discovered in the synthesized data.

% \subsubsection{Dataset}

The dataset used for clipper circuits modeling consisted of 7 minutes and 59 seconds of guitar and bass recordings from \cite{Abesser2013} and \cite{Kehling2014}, respectively. The amount of guitar recordings was roughly the same as the amount of bass recordings and their ordering was arbitrary. All recordings were single-channel and used \SI{44.1}{kHz} sampling rate. The target distortion signals were synthesized from SPICE models of the circuits with the schematics from \linebreak \Figures{fig:diode_clipper_circuit}{fig:diode2_clipper_circuit} using LTspice XVII by Analog Devices. Approximately 20\% of the dataset were used as the test set. The remaining data was split into the bass-only validation set and the training set according to the 80:20 rule.

The loss functions and dataset handling were implemented using the CoreAudioML library\footnote{\url{https://www.github.com/Alec-Wright/CoreAudioML}}. The remainder of the pipeline was implemented using the PyTorch library. We used exponential learning rates and the Adam optimizer \cite{Kingma2017}.

\subsection{First-Order Diode Clipper}
\label{subsec:diode_clipper1}

\begin{figure}
  \centering
  \scalebox{0.8}{\begin{tikzpicture}[straight voltages]
%--------start graphics code --------
% \draw[step=0.5,very thin, black!20] (-1,-0.5) grid (6,2.5);
\path (0,0) coordinate (ref_gnd);
\draw
  (ref_gnd)++(0,2) node[ocirc] {}
  node[xshift=-2mm,yshift=4mm] {$5 V_\text{in}$}
  to [R, l=$R{=}\SI{2.2}{\kilo\ohm}$] ++(2,0) node[circ] {}
  to ++(2,0)  node[circ] {}
  to ++(1,0)  node[circ] {}
  to ++(1,0)  node[ocirc] {} {}
  node[yshift=4mm] {$V_\text{out}$}
  ++(-1,0)
  to [D] ++(0,-2)
  to ++(-1,0) node[circ] {}
  to [D] ++(0,2)
  ++(0,-2)
  to ++(-1,0)
  node[ground] {}
  to ++(-1,0)
  to [C, l=$C{=}\SI{0.01}{\micro\farad}$, v=$y_1$] ++(0,2);
%--------end graphics code ----------
\end{tikzpicture}}
  \caption{First-order diode clipper circuit with marked state $y_1$. Input voltage $V_\text{in}$ was scaled by 5 in the target signal synthesis process.}
  \label{fig:diode_clipper_circuit}
\end{figure}
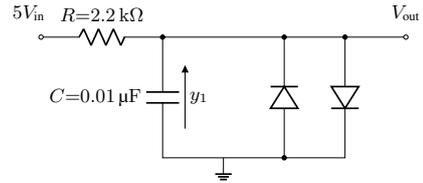

The first-order diode clipper is a circuit used to achieve signal distortion, e.g., in guitar effects pedals \cite{Yeh2007}. Its schematic is shown in \Figure{fig:diode_clipper_circuit}. It can be regarded as consisting of two parts: an RC low-pass filter and a diode limiter.

The first-order diode clipper is a system particularly interesting in the context of ODENet, because it is governed by a known, nonlinear \ac{ODE} derived by Yeh \cite{Yeh2007,Yeh2008}.
% JW: Since the first-order diode clipper equation is described by Yeh, it is pointless to include it here.
% \begin{equation}
%   \frac{\mathrm{d} V_\text{out}}{\mathrm{d}t} = \frac{V_\text{in} - V_\text{out}}{RC} - 2 \frac{I_\text{s}}{C} \sinh \left(\frac{V_\text{out}}{V_\text{t}}\right),
%   \label{eq:diode_clipper_equation}
% \end{equation}
% where $V_\text{in}$ is the input voltage (typically on the order of volts), $V_\text{out}$ is the output voltage, $t$ denotes time, $R$ is the serial resistance, $C$ is the parallel capacitance, $I_\text{s}$ is the reverse saturation current, and $V_\text{t}$ is the thermal voltage. The last two are parameters of the diodes that can be measured \cite{Yeh2007}. 
The circuit has one state (voltage across a capacitor), $y_1$, which is taken as the output voltage $V_\text{out}$. The first-order diode clipper had already been modeled with \ac{STN} \cite{Parker2019}.

\subsubsection{Compared Models}
 
The smallest derivative network that has reached a validation loss smaller than the assumed arbitrary threshold of 0.01 was a $2 \times 9 \times 9 \times 9 \times 1$ \ac{MLP} (ODENet9).
This network was subsequently trained in two setups: i)~with the forward Euler (FE) scheme and the \ac{ReLU} nonlinearity and ii)~with the implicit Adams-Bashforth-Moulton (IA) scheme \cite{Karlsson2019} and the \ac{SELU} nonlinearity \cite{klambauer2017selfnormalizing}.

For benchmark, we chose the \ac{STN} from \cite{Parker2019} of size $2 \times 4 \times 4 \times 4 \times 1$ with the tanh nonlinearity and bias enabled only in the second layer (STN $3 \times 4$ tanh), and a \ac{LSTM}-based architecture from \cite{Wright2019} with 8 hidden units and an $8 \times 1$ output \ac{MLP} mapping the hidden states to an output sample (\ac{LSTM}8). Each model is listed with its hyperparameters in \Table{tab:diode_clipper_models_data}. The number of epochs in training was determined by the early stopping criterion \cite{Goodfellow-et-al-2016}.

\begin{table}[t!]
    \centering
    \caption{Compared network architectures for diode clipper modeling.}
    % \resizebox{\textwidth}{!}{%
\begin{tabular}{@{}l c c @{}}
\toprule
\textbf{Model} & \textbf{\makecell{Number\\of\\parameters}} & \textbf{\makecell{Epochs\\in\\training}}\\ \midrule
LSTM8 & 361 & 200 \\ 
STN $3 \times 4$ tanh & 48 & 340 \\
ODENet9-FE ReLU & 127 & 300\\
ODENet9-IA SELU & 127 & 600\\ \bottomrule
\end{tabular}%
% }

    \label{tab:diode_clipper_models_data}
\end{table}

All of the architectures were trained on audio data at \SI{44.1}{kHz} sampling rate but tested on four different sampling rates to analyze the presence of aliasing in the output and inspect the interpretability of the learned \acp{ODE}. During the test, the \ac{STN} and ODENet models were informed about the new value of the time step, whereas for \ac{LSTM}8 it was not possible; \ac{LSTM} implicitly learns a fixed time step from the training data.

All models were tested on one long sequence, but the implicit Adams scheme consistently diverged in this test setting. Therefore, tests with ODENet9-IA were conducted using segments of \SI{22050}{} samples which were concatenated afterwards.

\subsubsection{Results}

In \Table{tab:diode_clipper_results}, test results of the compared models in terms of the \ac{SDR} rounded to one decimal place are shown. The best results (highest \ac{SDR}) are given in bold. The models were separately evaluated in terms of the learned model quality (the test sampling rate equal to the training sampling rate) and the performance at sampling rates different from the training sampling rate. 

\begin{table}[]
    \centering
    \caption{\Acl{SDR} across all test samples in dB for the first-order diode clipper models.}
    % % Loss value:
% \begin{tabular}{l c c c c}
%     \toprule
%     \textbf{\makecell{Test sampling\\rate [kHz]}} & \textbf{44.1} & \textbf{22.05} & \textbf{48} & \textbf{192} \\ \midrule
%     LSTM8 & \textbf{0.04\%} & \textbf{0.71\%} &  \textbf{0.04\%} & 0.55\% \\
%     STN 3x4 & 0.45\% & 230\% & 0.37\% & 0.36\% \\
%     ODENet9-FE &  0.37\% & 28\%  & 0.34\% & 0.22\% \\
%     ODENet9-IA & 0.11\% & 63\% & 0.09\% & \textbf{0.09\%} \\ \bottomrule
% \end{tabular}
% 
% % NMSE [dB]
% \begin{tabular}{l c c c c }
% \toprule
% \textbf{\makecell{Test sampling\\rate [kHz]}} & \textbf{44.1} & \textbf{22.05} & \textbf{48} & \textbf{192}
% \\ \midrule
% LSTM8 & \textbf{-30.9} & \textbf{-18.5} & \textbf{-30.7} & -19.6
% \\
% STN 3x4 & -20.4 & 6.6 & -21.3 & -21.4
% \\
% ODENet9-FE & -21.3 & -2.6 & -21.7 & -23.6
% \\
% ODENet9-IA & -26.4 & 0.8 & -27.3 & \textbf{-27.7}
% \\
% \bottomrule
% \end{tabular}

% SDR [dB]
\begin{tabular}{l c c c c }
\toprule
\textbf{\makecell{Test sampling\\rate [kHz]}} & \textbf{44.1} & \textbf{22.05} & \textbf{48} & \textbf{192}
\\ \midrule
LSTM8 & \textbf{30.9} & \textbf{18.5} & \textbf{30.7} & 19.6
\\
STN $3 \times 4$ & 20.4 & -6.6 & 21.3 & 21.5
\\
ODENet9-FE & 21.3 & 2.6 & 21.7 & 23.6
\\
ODENet9-IA & 26.4 & -0.8 & 27.3 & \textbf{27.7}
\\
\bottomrule
\end{tabular}
    \label{tab:diode_clipper_results}
\end{table}

%%%%%%%%%%%%%%%%%%%%%%%%%%%%%%%%%%%%%%%%%%%%%%%%%%%%%%%%%%%%%%%%%%%%%%%%%%%%%%%%%%%%%%
% \subsubsection{Model Quality}
%%%%%%%%%%%%%%%%%%%%%%%%%%%%%%%%%%%%%%%%%%%%%%%%%%%%%%%%%%%%%%%%%%%%%%%%%%%%%%%%%%%%%%%

At \SI{44.1}{kHz}, the \ac{SDR} was very high for all compared architectures (more than \SI{20}{dB}). In terms of the \ac{SDR}, the \ac{LSTM} outperformed all other models for \SI{22.05}{kHz}, \SI{44.1}{kHz}, and \SI{48}{kHz}.

%%%%%%%%%%%%%%%%%%%%%%%%%%%%%%%%%%%%%%%%%%%%%%%%%%%%%%%%%%%%%%%%%%%%%%%%%%%%%%%%%%%%%%%
% \subsubsection{Behavior at Unseen Sampling Rates}
%%%%%%%%%%%%%%%%%%%%%%%%%%%%%%%%%%%%%%%%%%%%%%%%%%%%%%%%%%%%%%%%%%%%%%%%%%%%%%%%%%%%%%%

\begin{figure}[t]
    \centering
    \includegraphics[width=\linewidth]{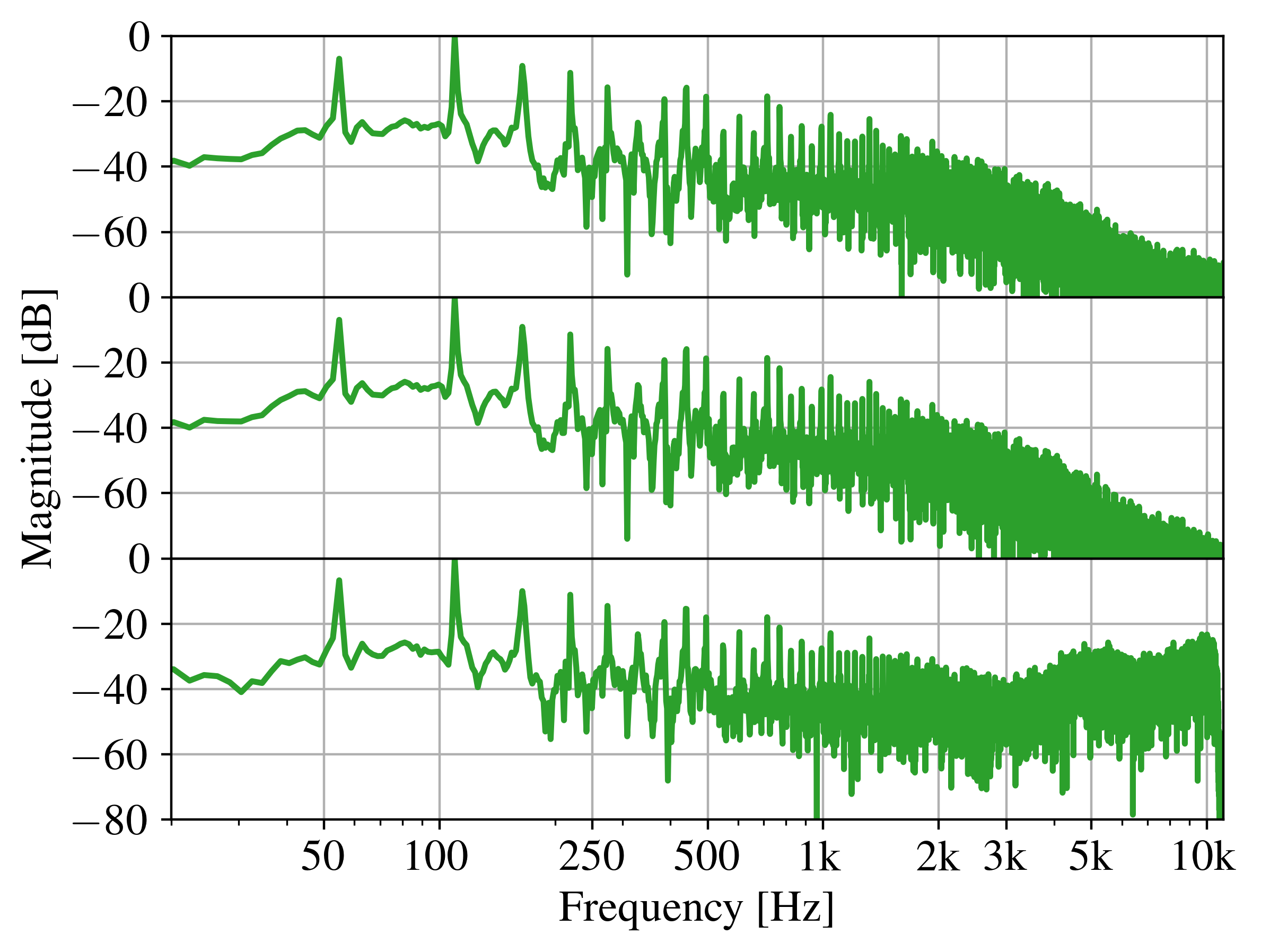}
    \caption{Magnitude spectrum of the A2 note played on the bass guitar as processed by the first-order diode clipper: (top) Target spectrum, (middle) ODENet9-FE output at \SI{192}{kHz} sampling rate, and (bottom) ODENet9-FE output at \SI{22.05}{kHz} sampling rate, showing aliasing at high frequencies. All spectra were normalized to the highest absolute value before the conversion to the decibel scale.}
    \label{fig:diode_clipper_aliasing}
\end{figure}

% For \SI{48000}{Hz} test sampling rate, again \ac{LSTM} outperformed other models in the analyzed metrics, yet all models had test loss value smaller than 1\%. In casual listening, there is not much difference between the model outputs.

\newcommand{\subfigureWidth}{\linewidth}

\begin{figure}[ht!]
    \centering
    \begin{subfigure}{\subfigureWidth}
        \centering
        % This file was created by tikzplotlib v0.9.9.
\begin{tikzpicture}

\begin{axis}[
colorbar horizontal,
colorbar style={at={(0,1.2)},anchor=north west,ytick={-50,-25,-7.105427357601e-15,25,50,75},yticklabels={
  \(\displaystyle {\ensuremath{-}50}\),
  \(\displaystyle {\ensuremath{-}25}\),
  \(\displaystyle {0}\),
  \(\displaystyle {25}\),
  \(\displaystyle {50}\),
  \(\displaystyle {75}\)
},ylabel={$10 \log_{10} |\frac{dy_1}{dt}|$}, ylabel style={at={(0,0.4)}}},
colormap/viridis,
point meta max=89.8890028498266,
point meta min=-59.9966700068274,
tick align=outside,
tick pos=left,
x grid style={white!69.0196078431373!black},
xlabel={$V_\text{in}$},
xmin=-1, xmax=1,
xtick style={color=black},
xtick={-1,-0.5,0,0.5,1},
xticklabels={
  \(\displaystyle {\ensuremath{-}1.0}\),
  \(\displaystyle {\ensuremath{-}0.5}\),
  \(\displaystyle {0.0}\),
  \(\displaystyle {0.5}\),
  \(\displaystyle {1.0}\)
},
y grid style={white!69.0196078431373!black},
ylabel={$y_1$},
ymin=-1, ymax=1,
ytick style={color=black},
ytick={-1,-0.5,0,0.5,1},
yticklabels={
  \(\displaystyle {\ensuremath{-}1.0}\),
  \(\displaystyle {\ensuremath{-}0.5}\),
  \(\displaystyle {0.0}\),
  \(\displaystyle {0.5}\),
  \(\displaystyle {1.0}\)
},
width=0.9\linewidth
]
\addplot graphics [includegraphics cmd=\pgfimage,xmin=-1, xmax=1, ymin=-1, ymax=1] {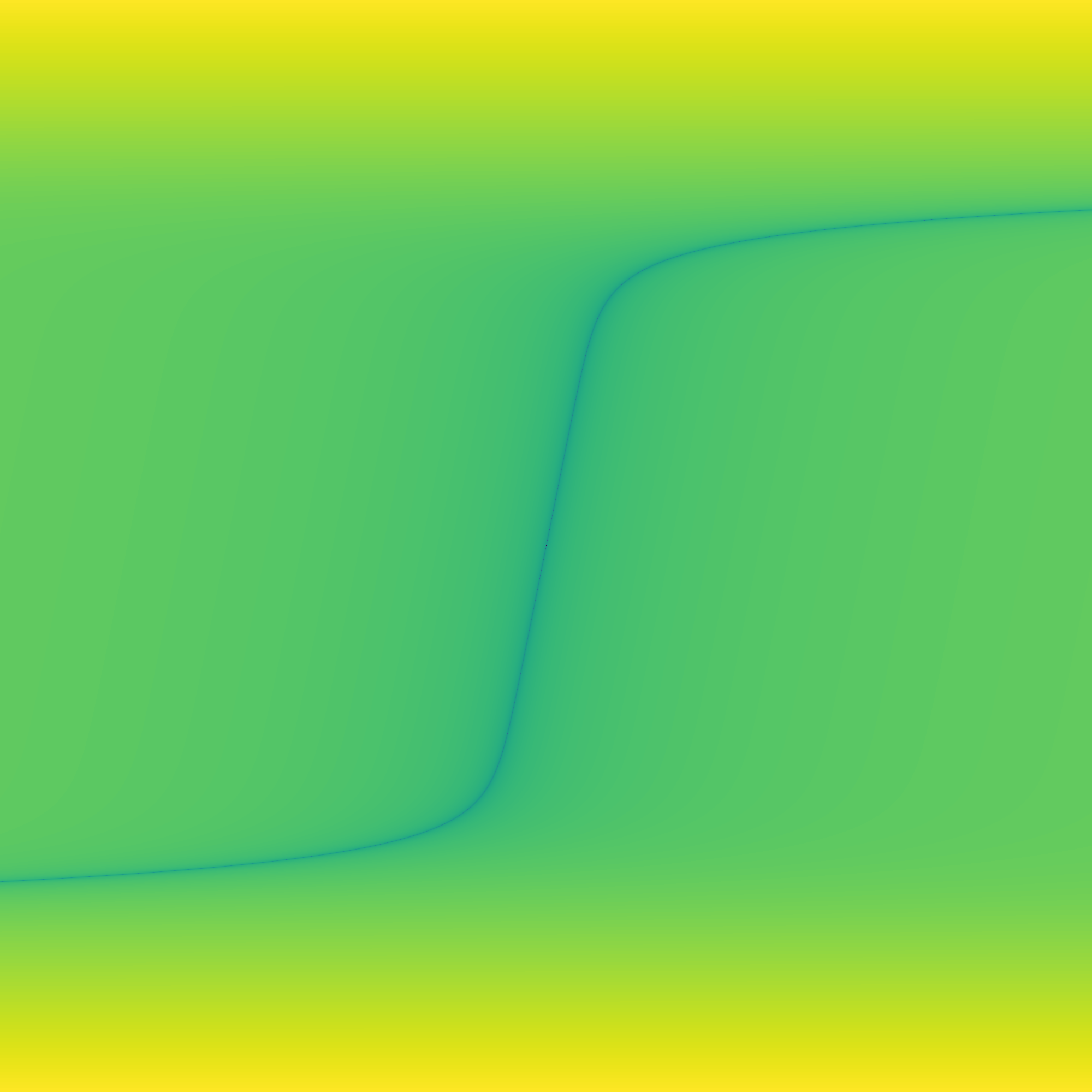};
\end{axis}

\end{tikzpicture}
        \caption{Closed-form. Every magnitude $m$ is transformed to $10 \log_{10}(m)$.}
    \end{subfigure}
    \begin{subfigure}{\subfigureWidth}
        \centering
        % This file was created by tikzplotlib v0.9.9.
\begin{tikzpicture}
\pgfplotsset{scaled ticks = false}
\begin{axis}[
colorbar horizontal,
colormap/viridis,
point meta max=106742.0625,
point meta min=0.010339617729187,
colorbar style={at={(0,1.2)},xtick={0,40000,80000},anchor=north west,ylabel={$|\frac{dy_1}{dt}|$}, ylabel style={at={(0,0.4)}}},
tick align=outside,
tick pos=left,
x grid style={white!69.0196078431373!black},
xlabel={$V_\text{in}$},
xmin=-1, xmax=1,
xtick style={color=black},
xtick={-1,-0.5,0,0.5,1},
xticklabels={
  \(\displaystyle {\ensuremath{-}1.0}\),
  \(\displaystyle {\ensuremath{-}0.5}\),
  \(\displaystyle {0.0}\),
  \(\displaystyle {0.5}\),
  \(\displaystyle {1.0}\)
},
y grid style={white!69.0196078431373!black},
ylabel={$y_1$},
ymin=-1, ymax=1,
ytick style={color=black},
ytick={-1,-0.5,0,0.5,1},
yticklabels={
  \(\displaystyle {\ensuremath{-}1.0}\),
  \(\displaystyle {\ensuremath{-}0.5}\),
  \(\displaystyle {0.0}\),
  \(\displaystyle {0.5}\),
  \(\displaystyle {1.0}\)
},
width=0.9\linewidth,
]
\addplot graphics [includegraphics cmd=\pgfimage,xmin=-1, xmax=1, ymin=-1, ymax=1] {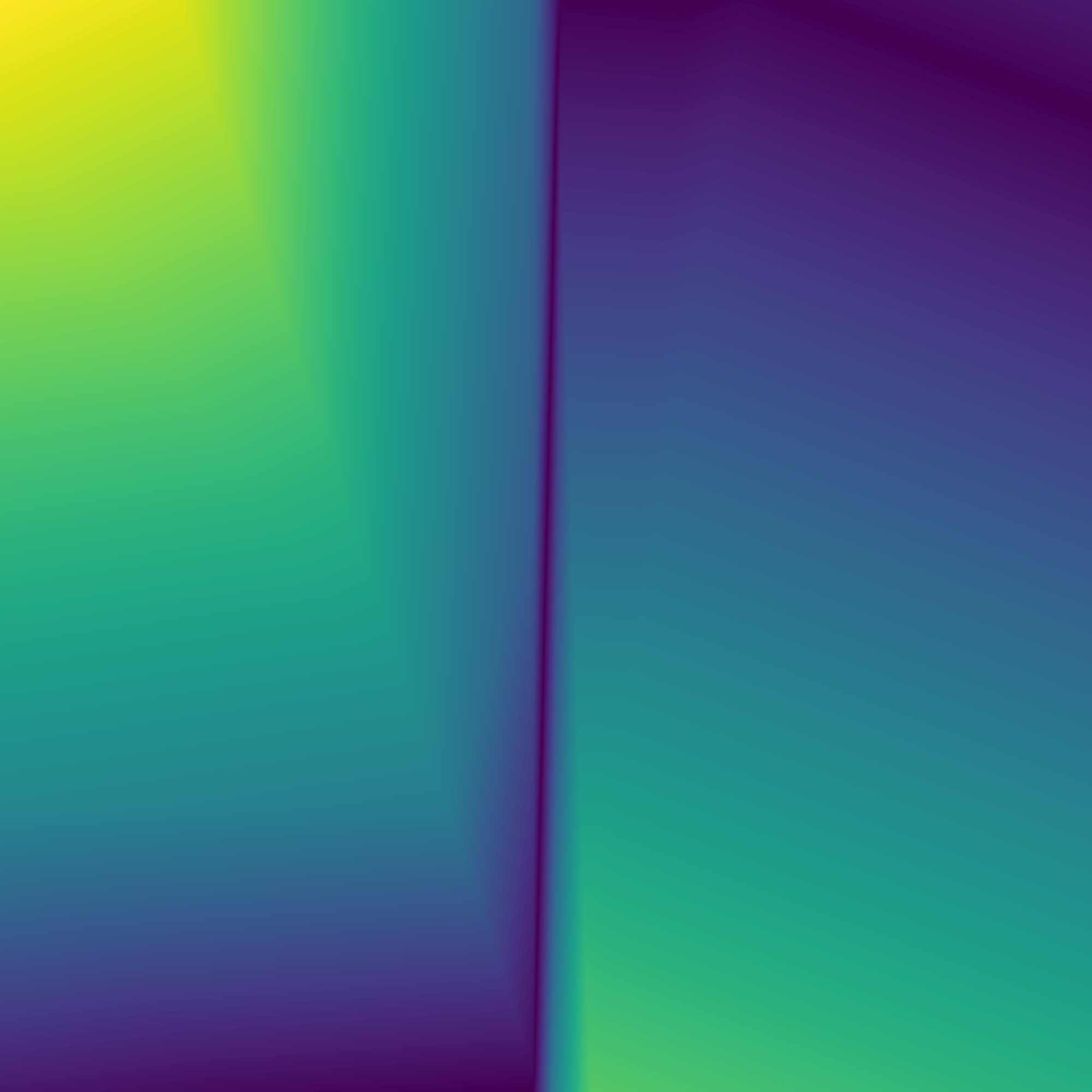};
\end{axis}

\end{tikzpicture}
        \caption{ODENet9-FE}
    \end{subfigure}
    % \begin{subfigure}{\subfigureWidth}
    %     \scalebox{0.76}{\input{figures/tikz/diode_clipper_derivative_visualization/IA-ode_derivative.tex}}
    %     \caption{ODENet9(IA)}
    % \end{subfigure}
    \caption{Magnitude of the derivative of the first-order diode clipper: analytical (top) and learned (bottom).}
    \label{fig:diode_clipper_derivative_visualizations}
\end{figure}

For \SI{22.05}{kHz} test sampling rate, \ac{STN} and ODENet models produced aliased output, which can be seen in \Figure{fig:diode_clipper_aliasing}. This is to be expected because these models were trained with a higher sampling rate.

\begin{figure}[h]
  \centering
  \begin{tikzpicture}[straight voltages]
%--------start graphics code --------
% \draw[step=0.5,very thin, black!20] (-1,-0.5) grid (6,2.5);
\path (0,0) coordinate (ref_gnd);
\draw
  (ref_gnd)++(0,2) node[ocirc] {}
  node[xshift=-2mm,yshift=4mm] {$5 V_\text{in}$}
  to [R, l=$R{=}$ \SI{2.2}{\kilo\ohm}] ++(2.5,0) {}
  to [C, l=$C_1{=}$ \SI{0.47}{\micro\farad}, v=$y_2$] ++(1,0) node[circ] {}
  to ++(1,0)  node[circ] {}
  to ++(1,0)  node[circ] {}
  to ++(0.2,0)  node[ocirc] {} {}
  node[yshift=4mm] {$V_\text{out}$}
  ++(-0.2,0)
  to [D] ++(0,-2)
  to ++(-1,0) node[circ] {}
  to [D] ++(0,2)
  ++(0,-2)
  to ++(-0.5,0)
  node[ground] {}
  to ++(-0.5,0)
  to [C, l=$C_2{=}$ \SI{0.01}{\micro\farad}, v=$y_1$] ++(0,2);
%--------end graphics code ----------
\end{tikzpicture}
  \caption{Second-order diode clipper circuit with marked states $y_1$ and $y_2$.}
  \label{fig:diode2_clipper_circuit}
\end{figure}
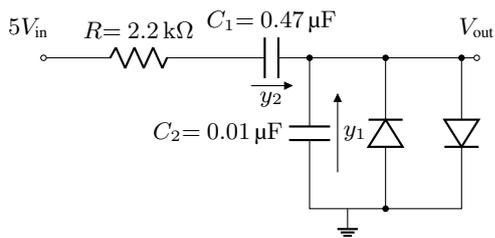

ODENet has some advantage for \SI{192}{kHz} test sampling rate, with both models having the two highest \ac{SDR} values. As could be expected, architectures that take the sampling rate into account during processing (\ac{STN}, ODENet) obtained a higher \ac{SDR} than the non-informed one. There was an advantage of the implicit solver over the explicit one in terms of the \ac{SDR} but at the cost of doubled time needed to process the same amount of data.

In \Figure{fig:diode_clipper_derivative_visualizations}, one can see the analytical derivative function from \cite{Yeh2007,Yeh2008} %\Equation{eq:diode_clipper_equation} 
and the one learned by ODENet9-FE. The learned derivative function, although similar in the S-shape, is much smoother than the analytical form. This difference probably comes from the limited frequency bandwidth of the dataset due to sampling. The learned derivative network visualization is on a par with the one shown in \cite{Parker2019}.

All in all, results comparable to the established \ac{LSTM} and \ac{STN} architectures prove that ODENet can be used for \ac{VA} modeling and at sampling rates higher than the training sampling rate even for the simplest numerical scheme, i.e., the forward Euler.

\subsection{Second-Order Diode Clipper}
\label{subsec:diode_clipper2}

The second-order diode clipper circuit and the values of the components were taken from \cite{Parker2019}. The circuit is similar to the first-order diode clipper with the only difference being an additional capacitor inserted between the output node and the resistor as shown in \Figure{fig:diode2_clipper_circuit}. This additional capacitor introduces high-pass filtering and makes the clipping asymmetrical \cite{Parker2019}.  The model has two states, $y_1$ and $y_2$. The former is taken as the circuit output voltage $V_\text{out}$. This circuit was modeled with \ac{STN} using real-world measurements in \cite{Parker2019}. However, we again used a SPICE simulation that processed the same dataset as in the case of the first-order diode clipper.

\subsubsection{Compared Models}

In \Table{tab:diode2_clipper_models_data}, the hyperparameters of the trained networks are presented. As a benchmark, we chose the \ac{LSTM}-based architecture with 16 memory cells (\acs{LSTM}16) and a $3 \times 30 \times 30 \times 2$  \ac{STN} (\ac{STN} $2 \times 30$). The derivative network of the ODENet was parametrized by a $3 \times 30 \times 30 \times 2$ \ac{MLP} (ODENet30) and a $3 \times 20 \times 20 \times 2$ \ac{MLP} (ODENet20), both with the Softsign nonlinearity. The former was chosen to have roughly the same number of trainable parameters as the benchmark, the latter was meant to be significantly smaller. The derivative network was used with the forward Euler (FE), trapezoidal rule (TR), and explicit Runge-Kutta of order 4 (RK4) numerical schemes \cite{Gockenbach2011}. The training proceeded as in the case of the first-order diode clipper.

\begin{table}[t!]
    \centering
    \caption{Compared network architectures for the second-order diode clipper modeling}
    % \resizebox{\textwidth}{!}{%
\begin{tabular}{@{}l c c @{}}
\toprule
\textbf{Model} & \textbf{\makecell{Number\\of\\parameters}} & \textbf{\makecell{Epochs\\in\\training}}\\ \midrule
LSTM16 & 1250 & 2000 \\ 
% LSTM32 & 4546 & 2000 \\ 
STN $2 \times 30$ tanh & 1112 & 630 \\
% ODENet10(FE) Softsign & 172 & 1200\\
ODENet20 Softsign & 542 & 1200\\
ODENet30 Softsign & 1112 & 1200\\ \bottomrule
% ODENet40(FE) Softsign & 1882 & 1200\\
\end{tabular}%
% }

    \label{tab:diode2_clipper_models_data}
\end{table}

\subsubsection{Results}

Results in terms of the \ac{SDR} are summarized in \Table{tab:diode2_clipper_results}. As in the case of the first-order diode clipper, the \acs{LSTM}-based architecture obtained the highest \ac{SDR} for \SI{44.1}{kHz} and \SI{48}{kHz}. Somewhat surprisingly, \ac{STN} obtained the highest \ac{SDR} for the remaining sampling rates, even for \SI{22.05}{kHz}, which introduced aliasing in this model's output for the first-order diode clipper. Again, the \ac{LSTM}-based architecture was inferior to other (sampling rate-informed) models at \SI{192}{kHz}.

ODENet models clearly performed worse, with the most sophisticated ODENet30-RK4 being close behind \ac{STN}. Results reveal, however, the expected outcome: more complicated solvers perform better than simple ones even though they are using the same derivative network architecture. Additionally, a smaller network using an advanced scheme ODENet20-RK4 outperformed a larger network using a simpler scheme ODENet30-FE, but this result should be treated more as a hint rather than a general rule.

% and \SI{22050}{Hz}, where the former was the sampling rate of the training set. However, the difference between \acs{LSTM}16 and ODENet30 for \SI{44100}{Hz} is not significant. On the other hand, already for \SI{48000}{Hz}, ODENet30 obtained the lowest loss and for \SI{192000}{Hz}, significantly outperformed \acs{LSTM}16. In the last scenario even the two times smaller ODENet20 obtained a lower loss than \acs{LSTM}16. What is more, ODENet30 improved its loss with increasing sampling rate.

\begin{figure}
    \centering
    \input{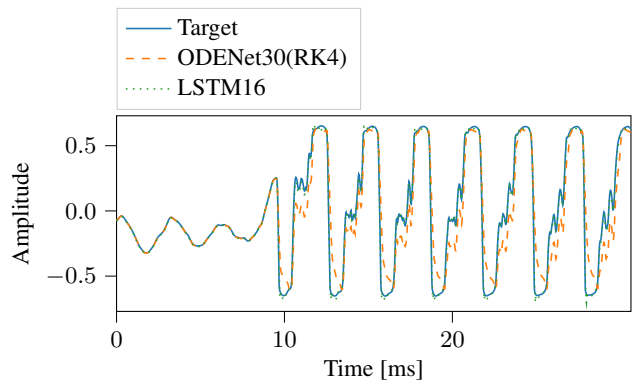}
    \caption{Time-domain comparison of the results of the second-order diode clipper modeling. Presented signal is a guitar note.}
    \label{fig:diode2_time_alignment}
\end{figure}

\Figure{fig:diode2_time_alignment} shows the accuracy of \acs{LSTM}16 and ODENet30-RK4 at \SI{44.1}{kHz} in the time domain. ODENet seems to follow the signal well for small oscillations but fails to match the peaks of the waveform.

\begin{table}[]
    \centering
    \caption{\Acl{SDR} across all test samples in dB for the second-order diode clipper models.}
    \label{tab:diode2_clipper_results}
\begin{tabular}{l c c c c }
\toprule
\textbf{\makecell{Test sampling\\rate [kHz]}} & \textbf{44.1} & \textbf{22.05} & \textbf{48} & \textbf{192}
\\ \midrule
LSTM16 & \textbf{21.2} & 13.5 & \textbf{20.6} & 7.9
\\
STN $2 \times 30$ & 16.9 & \textbf{15.7} & 16.8 & \textbf{16.3}
\\
ODENet20-RK4  & 14.3 & 9.0 & 14.2 & 14.1
\\
ODENet30-FE & 11.6 & 6.4 & 11.5 & 11.0
\\
ODENet30-TR & 15.5 & 8.2 & 15.4 & 13.9
\\
ODENet30-RK4 & 15.7 & 14.7 & 15.6 & 15.7
\\
\bottomrule
\end{tabular}
\end{table}

% \begin{figure}
%     \centering
%     \includegraphics[width=0.9\linewidth,height=4cm]{figures/frequency_alignment_comparison.png}
%     \caption{Frequency-domain comparison of the results of the second-order diode clipper. Presented spectra are of the A2 note played on a bass guitar and lie in the \SIrange{2}{3}{\kilo\Hz} range.}
%     \label{fig:diode2_frequency_alignment}
% \end{figure}

\begin{figure}[t]
    \centering
    \begin{subfigure}{\linewidth}
        \centering
        % This file was created by tikzplotlib v0.9.6.
\begin{tikzpicture}

\begin{axis}[
colorbar horizontal,
colorbar style={at={(0,1.2)},anchor=north west,ylabel={$|\frac{dy_1}{dt}|$}, ylabel style={at={(0.05,-0.4)},anchor=north west}},
colormap/viridis,
point meta max=3336.61059570312,
point meta min=1.52587890625e-05,
tick align=outside,
tick pos=left,
x grid style={white!69.0196078431373!black},
xlabel={$V_\text{in}$},
xmin=-1, xmax=1,
xtick style={color=black},
xtick={-1,-0.5,0,0.5,1},
xticklabels={\(\displaystyle {\ensuremath{-}1.0}\),\(\displaystyle {\ensuremath{-}0.5}\),\(\displaystyle {0.0}\),\(\displaystyle {0.5}\),\(\displaystyle {1.0}\)},
y grid style={white!69.0196078431373!black},
ylabel={$y_1$},
ymin=-1, ymax=1,
ytick style={color=black},
ytick={-1,-0.75,-0.5,-0.25,0,0.25,0.5,0.75,1},
yticklabels={\(\displaystyle {\ensuremath{-}1.00}\),\(\displaystyle {\ensuremath{-}0.75}\),\(\displaystyle {\ensuremath{-}0.50}\),\(\displaystyle {\ensuremath{-}0.25}\),\(\displaystyle {0.00}\),\(\displaystyle {0.25}\),\(\displaystyle {0.50}\),\(\displaystyle {0.75}\),\(\displaystyle {1.00}\)}, 
width=0.9\linewidth,
]
\addplot graphics [includegraphics cmd=\pgfimage,xmin=-1, xmax=1, ymin=-1, ymax=1] {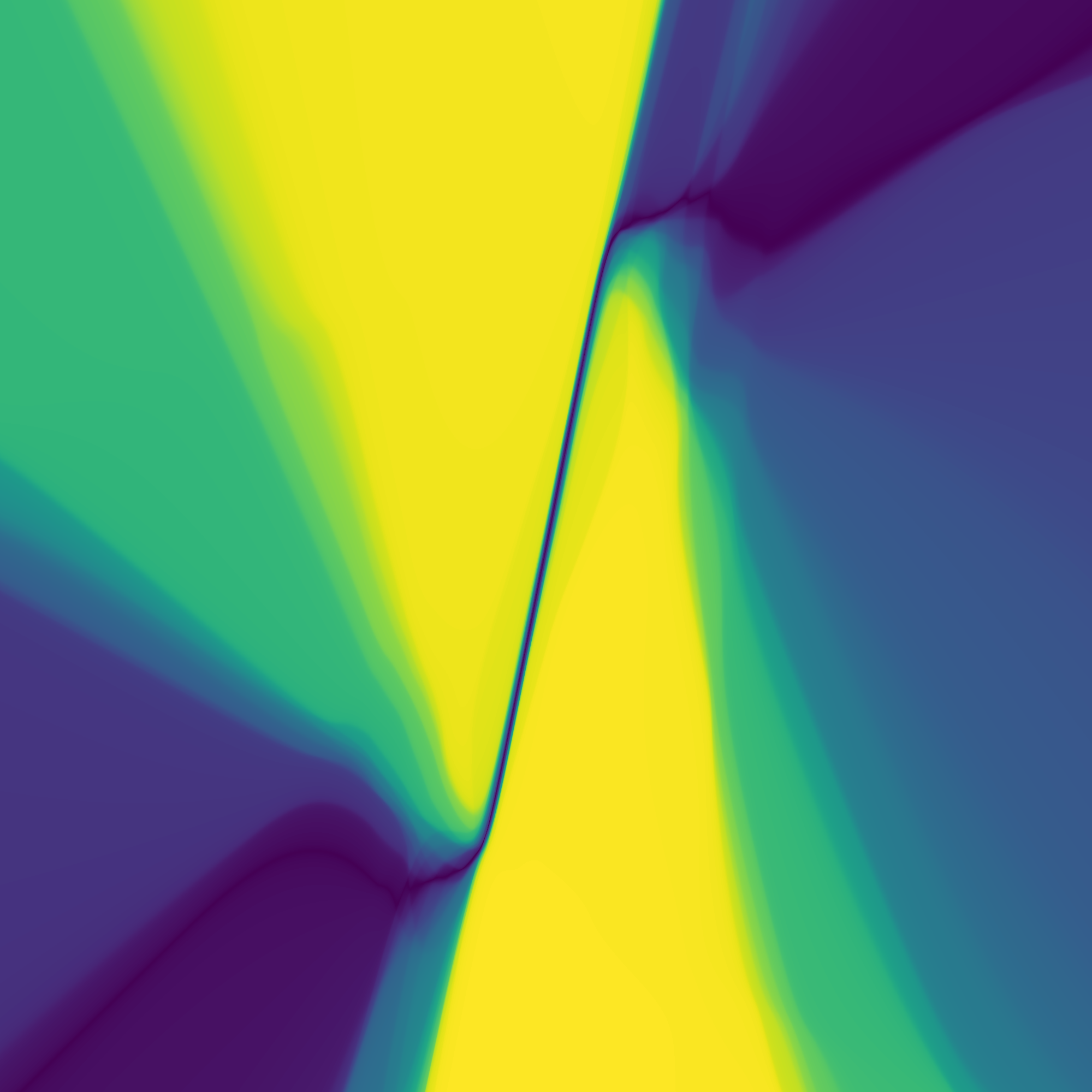};
\end{axis}

\end{tikzpicture}
        \caption{$y_2=0.$}
        \label{fig:state1_derivative_y2_0}
    \end{subfigure}
    \begin{subfigure}{\linewidth}
        \centering
        % This file was created by tikzplotlib v0.9.6.
\begin{tikzpicture}

\begin{axis}[
% colorbar horizontal,
% colorbar style={at={(0,1.2)},anchor=north west,ytick={500,1000,1500,2000,2500,3000},yticklabels={\(\displaystyle {500}\),\(\displaystyle {1000}\),\(\displaystyle {1500}\),\(\displaystyle {2000}\),\(\displaystyle {2500}\),\(\displaystyle {3000}\)},ylabel={$|\frac{dy_1}{dt}|$}, ylabel style={at={(0.05,-0.4)},anchor=north west}},
% colormap/viridis,
point meta max=3332.01000976562,
point meta min=0.0001373291015625,
tick align=outside,
tick pos=left,
x grid style={white!69.0196078431373!black},
xlabel={$V_\text{in}$},
xmin=-1, xmax=1,
xtick style={color=black},
xtick={-1,-0.5,0,0.5,1},
xticklabels={\(\displaystyle {\ensuremath{-}1.0}\),\(\displaystyle {\ensuremath{-}0.5}\),\(\displaystyle {0.0}\),\(\displaystyle {0.5}\),\(\displaystyle {1.0}\)},
y grid style={white!69.0196078431373!black},
ylabel={$y_1$},
ymin=-1, ymax=1,
ytick style={color=black},
ytick={-1,-0.75,-0.5,-0.25,0,0.25,0.5,0.75,1},
yticklabels={\(\displaystyle {\ensuremath{-}1.00}\),\(\displaystyle {\ensuremath{-}0.75}\),\(\displaystyle {\ensuremath{-}0.50}\),\(\displaystyle {\ensuremath{-}0.25}\),\(\displaystyle {0.00}\),\(\displaystyle {0.25}\),\(\displaystyle {0.50}\),\(\displaystyle {0.75}\),\(\displaystyle {1.00}\)}, 
width=0.9\linewidth,
]
\addplot graphics [includegraphics cmd=\pgfimage,xmin=-1, xmax=1, ymin=-1, ymax=1] {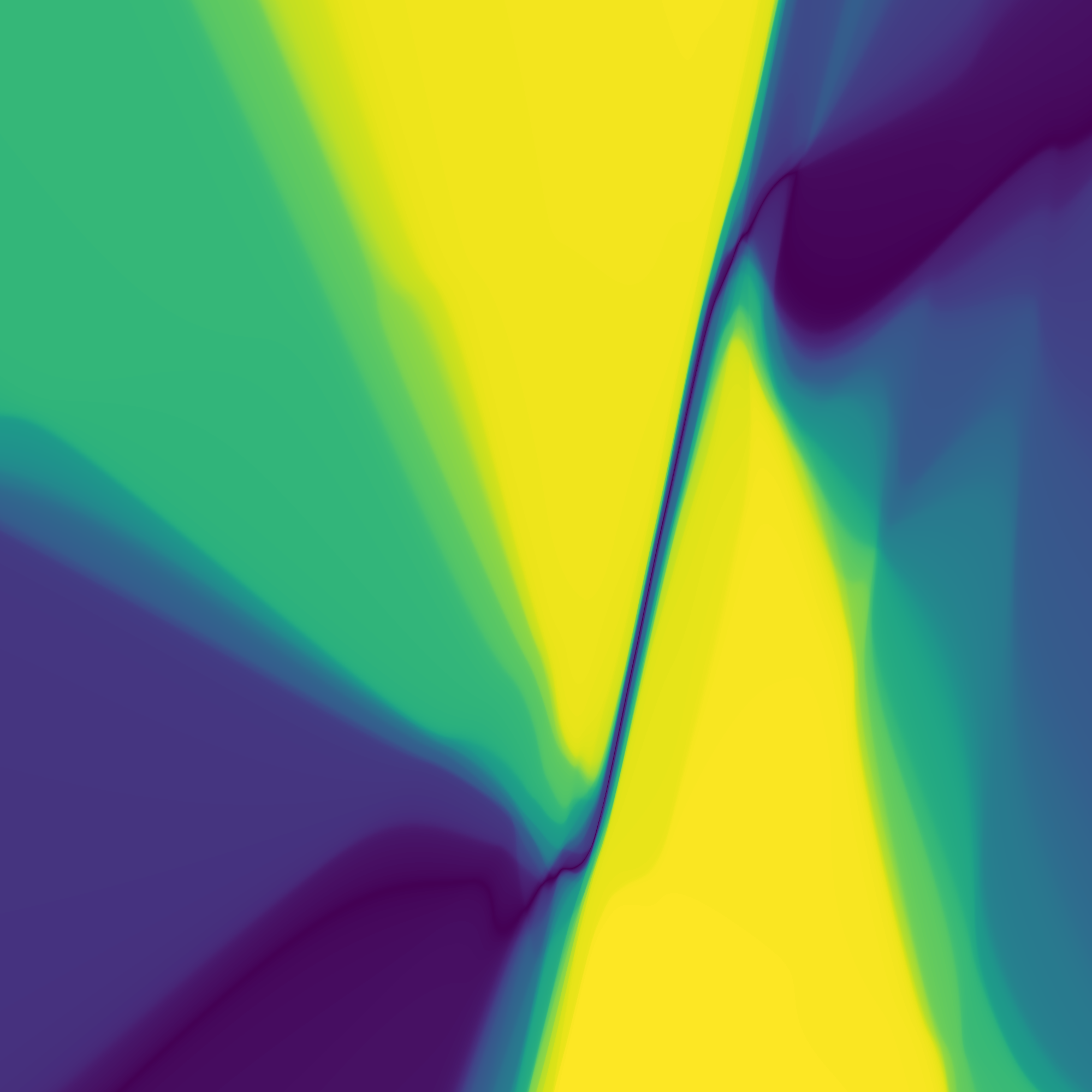};
\end{axis}

\end{tikzpicture}
        \caption{$y_2=1.$}
        \label{fig:state1_derivative_y2_1}
    \end{subfigure}
    \caption{Magnitude of the learned derivative of the first state of the second-order diode clipper for two different values of the second state.}
    \label{fig:state1_derivative}
\end{figure}

\begin{figure}
    \centering
    % This file was created by tikzplotlib v0.9.6.
\begin{tikzpicture}

\begin{axis}[
colorbar horizontal,
colorbar style={at={(0,1.2)},anchor=north west,ylabel={$|\frac{dy_2}{dt}|$}, ylabel style={at={(0.05,-0.4)},anchor=north west}},
colormap/viridis,
point meta max=684.078735351562,
point meta min=0.000400543212890625,
tick align=outside,
tick pos=left,
x grid style={white!69.0196078431373!black},
xlabel={$V_\text{in}$},
xmin=-1, xmax=1,
xtick style={color=black},
xtick={-1,-0.5,0,0.5,1},
xticklabels={\(\displaystyle {\ensuremath{-}1.0}\),\(\displaystyle {\ensuremath{-}0.5}\),\(\displaystyle {0.0}\),\(\displaystyle {0.5}\),\(\displaystyle {1.0}\)},
y grid style={white!69.0196078431373!black},
ylabel={$y_1$},
ymin=-1, ymax=1,
ytick style={color=black},
ytick={-1,-0.75,-0.5,-0.25,0,0.25,0.5,0.75,1},
yticklabels={\(\displaystyle {\ensuremath{-}1.00}\),\(\displaystyle {\ensuremath{-}0.75}\),\(\displaystyle {\ensuremath{-}0.50}\),\(\displaystyle {\ensuremath{-}0.25}\),\(\displaystyle {0.00}\),\(\displaystyle {0.25}\),\(\displaystyle {0.50}\),\(\displaystyle {0.75}\),\(\displaystyle {1.00}\)}, 
width=0.9\linewidth,
]
\addplot graphics [includegraphics cmd=\pgfimage,xmin=-1, xmax=1, ymin=-1, ymax=1] {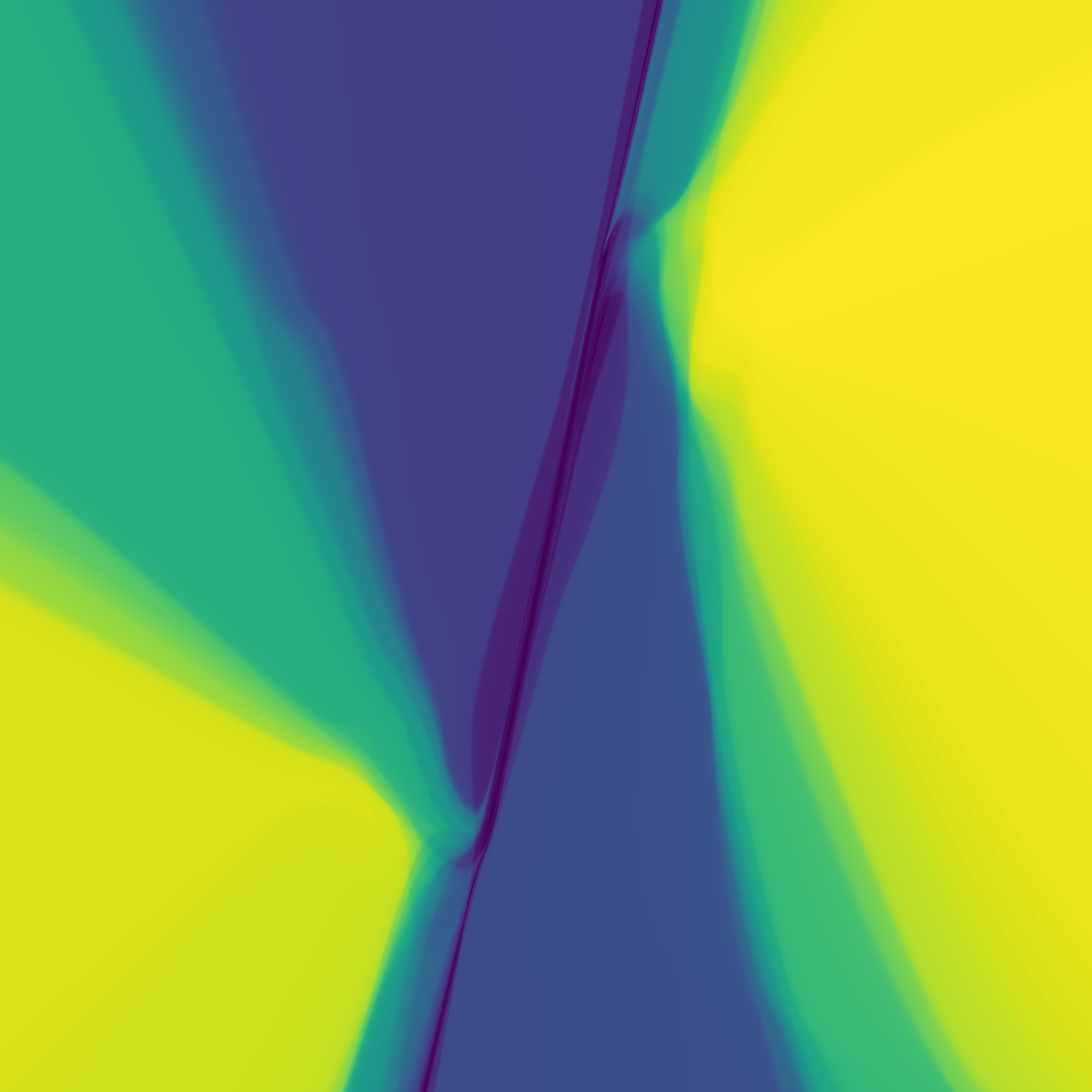};
\end{axis}

\end{tikzpicture}
    \caption{Magnitude of the learned derivative of the second state of the second-order diode clipper for $y_2 = 0$.}
    \label{fig:state2_derivative}
\end{figure}

One can obtain more insight into the ODENet by inspecting the learned derivative of the state. The derivative is a vector-valued function of three variables: the input voltage, $V_\text{in}$, and the two states, $y_1$ and $y_2$. In \Figure{fig:state1_derivative} the magnitude of the derivative of the first state learned by ODENet30-RK4 is shown for two values of $y_2$. One can imagine these figures as snapshots taken at different positions of the state space. The magnitude of the derivative of the second state for $y_2=0$ can be seen in \Figure{fig:state2_derivative}. If the magnitude of the learned derivative is asymmetrical with respect to the input voltage, the chosen value for $y_2$ should make it clear.

The derivative of the first state resembles the S-shape of the first-order diode clipper derivative. The clipping behavior manifests itself in regions with large derivative magnitudes, which ``push'' the output towards the S-shape. Furthermore, for $y_2 = 1$, the behavior of the clipper becomes asymmetrical, which corresponds to the previous analysis of the circuit in \cite{Parker2019}. One can see it even better in \Figure{fig:state2_derivative}, which shows that the derivative of the second state is inherently asymmetrical with respect to the input or $y_1$ state. As one can see from \Figures{fig:state1_derivative}{fig:state2_derivative}, the learned derivative would be challenging to derive analytically in a white-box fashion, which confirms the usefulness of the ODENet in \ac{VA} modeling.

%%%%%%%%%%%%%%%%%%%%%%%%%%%%%%%%%%%%%%%%%%%%%%%%%%%%%%%%%
\subsection{Time-Varying Models}
\label{subsubsec:time_varying_modeling}
%%%%%%%%%%%%%%%%%%%%%%%%%%%%%%%%%%%%%%%%%%%%%%%%%%%%%%%%%

We also applied ODENet to phaser modeling following the ``toy-problem'' approach from \cite{Wright2020}. However, even with ground truth-informed dataset, ODENet was severely underfitting contrary to the baseline from \cite{Wright2020}. There seems to be an inherent difficulty for ODENet in learning this time-varying system. This observation was confirmed in \cite{Peussaetal2021}, where \ac{STN}, which is similar in nature to ODENet, failed to fit a phaser dataset. Possible explanations and solutions of this difficulty could be investigated in future work.

%%%%%%%%%%%%%%%%%%%%%%%%%%%%%%%%%%%%%%%%%%%%%%%%%%%%%%%%%
\subsection{Challenges}
\label{subsec:discussion}
%%%%%%%%%%%%%%%%%%%%%%%%%%%%%%%%%%%%%%%%%%%%%%%%%%%%%%%%%

Applying ODENet to \ac{VA} is not without challenges. One needs to obtain the state data to successfully train the derivative network. To know which state data to capture, some insight into the modeled device is needed as in white-box modeling. Additionally, the implementation we used was quite slow to train because ODENet allows parallelization only through the usage of minibatches.

\section{Conclusions}
\label{sec:conclusions}

In this paper, the concept of learning \acp{ODE} from data and solving them numerically for \ac{VA} modeling was presented. We adapted this framework, termed ODENet, to handle an input signal and the initial conditions of \acp{ODE} describing analog electronic circuits. We successfully applied it to \ac{VA} modeling of two distortion circuits: the first- and the second-order diode clipper. Our approach obtained comparable performance to the baseline at the sampling rate of the training set while using fewer trainable parameters and showed acceptable performance at increased sampling rates. In some cases, the accuracy increased at higher sampling rates. The learned derivative is physically interpretable but does not have to be derived analytically and does not need oversampling, which are big advantages over purely white-box models. The learned derivative may be used in conjunction with any numerical solver, which allows an accuracy-performance trade-off. Future work could include modeling systems with more states or time-varying systems, or a deeper analysis of the learned \ac{ODE}-numerical solver coupling.

\section{Acknowledgments}

The main part of the work was conducted in July--November 2021, when the first author was visiting the Aalto Acoustics Lab in Espoo, Finland. He is deeply thankful to the members of the lab for inspiring conversations and the feedback related to this work. The authors also thank anonymous reviewers for their insightful comments.

%\newpage
% \nocite{*} commented by J.W.
\bibliographystyle{IEEEbib}
\bibliography{DAFx22_tmpl} % requires file DAFx20_tmpl.bib

\begin{thebibliography}{10}

\bibitem{Zoelzer2011}
U.~Z{\"o}lzer,
\newblock {\em {DAFX}: {D}igital {A}udio {E}ffects},
\newblock John Wiley \& Sons, Ltd, 2nd edition, 2011.

\bibitem{Parker2019}
J.~D. Parker, F.~Esqueda, and A.~Bergner,
\newblock ``Modelling of {N}onlinear {S}tate-{S}pace {S}ystems {U}sing a {D}eep
  {N}eural {N}etwork,''
\newblock in {\em Proc. of the 22nd Int. Conf. on Digital Audio Effects (DAFx),
  Birmingham, UK}, 2019.

\bibitem{Eichas2014}
F.~Eichas, M.~Fink, M.~Holters, and U.~Zölzer,
\newblock ``Physical {M}odeling of the {MXR} {P}hase 90 {G}uitar {E}ffect
  {P}edal,''
\newblock in {\em Proc. of the 17th Int. Conf. on Digital Audio Effects (DAFx),
  Erlangen, Germany, September 1-5}, 2014.

\bibitem{Kiiski2016}
R.~Kiiski, F.~Esqueda, and V.~Välimäki,
\newblock ``Time-{V}ariant {G}ray-{B}ox {M}odeling of a {P}haser {P}edal,''
\newblock in {\em Proc. of the 19th Int. Conf. on Digital Audio Effects (DAFx),
  Brno, Czech Republic, September 5–9}, 2016, pp. 121--128.

\bibitem{Wright2020}
A.~Wright and V.~Välimäki,
\newblock ``Neural {M}odelling of {P}eriodically {M}odulated {T}ime-{V}arying
  {E}ffects,''
\newblock in {\em Proc. of the 23rd Int. Conf. on Digital Audio Effects (DAFx),
  Vienna, Austria, September 2020-21}, 2020.

\bibitem{Wright2019a}
A.~Wright and V.~Välimäki,
\newblock ``Perceptual {L}oss {F}unction for {N}eural {M}odelling of {A}udio
  {S}ystems,''
\newblock in {\em Proc. of the IEEE Int. Conf. on Acoustics, Speech and Signal
  Processing (ICASSP)}, May 2020.

\bibitem{Wright2019}
A.~Wright, E.-P. Damskägg, and V.~Välimäki,
\newblock ``Real-{T}ime {B}lack-{B}ox {M}odelling with {R}ecurrent {N}eural
  {N}etworks,''
\newblock in {\em Proc. of the 22nd Int. Conf. on Digital Audio Effects
  (DAFx-19), Birmingham, UK}, 2019.

\bibitem{Wrightetal2020}
A.~Wright, E.-P. Damskägg, L.~Juvela, and V.~Välimäki,
\newblock ``Real-{T}ime {G}uitar {A}mplifier {E}mulation with {D}eep
  {L}earning,''
\newblock {\em Applied Sciences}, vol. 10, no. 3, 2020.

\bibitem{Parker2013}
J.~D. Parker,
\newblock {\em Dispersive {S}ystems in {M}usical {A}udio {S}ignal
  {P}rocessing},
\newblock Ph.D. thesis, Aalto University, Espoo, Finland, Oct. 2013.

\bibitem{Pakarinenetal2011}
J.~Pakarinen, V.~Välimäki, F.~Fontana, V.~Lazzarini, and J.~Abel,
\newblock ``Recent {A}dvances in {R}eal-{T}ime {M}usical {E}ffects,
  {S}ynthesis, and {V}irtual {A}nalog {M}odels,''
\newblock {\em EURASIP J. Adv. Sig. Proc.}, vol. 2011, January 2011.

\bibitem{Yeh2007}
D.~Yeh, J.~Abel, and J.~Smith,
\newblock ``Simulation of the {D}iode {L}imiter in {G}uitar {D}istortion
  {C}ircuits by {N}umerical {S}olution of {O}rdinary {D}ifferential
  {E}quations,''
\newblock in {\em Proc. of the 10th Int. Conf. on Digital Audio Effects (DAFx),
  Bordeaux, France, September 10-15}, 2007.

\bibitem{Yeh2008}
D.~T. Yeh, J.~S. Abel, A.~Vladimirescu, and J.~O. Smith,
\newblock ``Numerical {M}ethods for {S}imulation of {G}uitar {D}istortion
  {C}ircuits,''
\newblock {\em Computer Music Journal}, vol. 32, no. 2, pp. 23--42, 2008.

\bibitem{PASPWEB2010}
J.~O. Smith,
\newblock {\em Physical {A}udio {S}ignal {P}rocessing},
\newblock
  \href{https://ccrma.stanford.edu/~jos/pasp/}{http://ccrma.stanford.edu/$_{\widetilde{~}}$jos/pasp/},
  2010,
\newblock Online book, 2010 edition. Retrieved June 18, 2021.

\bibitem{Esquedaetal2021}
F.~Esqueda, B.~Kuznetsov, and J.~D. Parker,
\newblock ``Differentiable {W}hite-{B}ox {V}irtual {A}nalog {M}odeling,''
\newblock in {\em Proc. of the 23rd Int. Conf. on Digital Audio Effects (DAFx),
  Vienna, Austria, September 2020-21}, 2021.

\bibitem{He2015}
K.~He, X.~Zhang, S.~Ren, and J.~Sun,
\newblock ``Deep {R}esidual {L}earning for {I}mage {R}ecognition,''
\newblock {\em Proc. IEEE Conf. on Computer Vision and Pattern Recognition
  (CVPR)}, pp. 770--778, 2016.

\bibitem{Chen2018}
R.~T.~Q. Chen, Y.~Rubanova, J.~Bettencourt, and D.~Duvenaud,
\newblock ``Neural {O}rdinary {D}ifferential {E}quations,''
\newblock in {\em Proc. of the 31st Conf. on Neural Information Processing
  Systems (NeurIPS 2018), Montréal, Canada}, 2018.

\bibitem{Karlsson2019}
D.~Karlsson and O.~Svanstr\"om,
\newblock ``Modelling {D}ynamical {S}ystems {U}sing {N}eural {O}rdinary
  {D}ifferential {E}quations,''
\newblock M.S. thesis, Chalmers University of Technology, 2019.

\bibitem{Gockenbach2011}
M.~Gockenbach,
\newblock {\em Partial {D}ifferential {E}quations: {A}nalytical and {N}umerical
  {M}ethods},
\newblock Society for Industrial and Applied Mathematics, Philadelphia, 2011.

\bibitem{Goodfellow-et-al-2016}
I.~Goodfellow, Y.~Bengio, and A.~Courville,
\newblock {\em Deep {L}earning},
\newblock MIT Press, 2016,
\newblock
  \href{http://www.deeplearningbook.org}{http://www.deeplearningbook.org}.

\bibitem{Peussaetal2021}
A.~Peussa, E.-P. Damskägg, S.~Mimilakis, T.~Sherson, L.~Juvela,
  A.~Gotsopoulos, and V.~Välimäki,
\newblock ``Exposure {B}ias and {S}tate {M}atching in {R}ecurrent {N}eural
  {N}etwork {V}irtual {A}nalog {M}odels,''
\newblock in {\em Proc. of the 23rd Int. Conf. on Digital Audio Effects (DAFx),
  Vienna, Austria, September 2020-21}, 2021.

\bibitem{Smith2018}
L.~N. Smith,
\newblock ``A {D}isciplined {A}pproach to {N}eural {N}etwork
  {H}yper-parameters: {P}art 1 -- {L}earning {R}ate, {B}atch {S}ize,
  {M}omentum, and {W}eight {D}ecay,''
\newblock {\em US Naval Research Laboratory Technical Report 5510-026}, 2018.

\bibitem{Abesser2013}
J.~Abeßer, P.~Kramer, Ch. Dittmar, and G.~Schuller,
\newblock ``Parametric {A}udio {C}oding of {B}ass {G}uitar {R}ecordings using a
  {T}uned {P}hysical {M}odeling {A}lgorithm,''
\newblock in {\em Proc. of the 16th Int. Conf. on Digital Audio Effects (DAFx),
  Maynooth, Ireland, September 2-5}, 2013.

\bibitem{Kehling2014}
Ch. Kehling, J.~Abeßer, Ch. Dittmar, and G.~Schuller,
\newblock ``Automatic {T}ablature {T}ranscription of {E}lectric {G}uitar
  {R}ecordings by {E}stimation of {S}core- and {I}nstrument-related
  {P}arameters,''
\newblock in {\em Proc. of the 17th Int. Conf. on Digital Audio Effects (DAFx),
  Erlangen, Germany, September 1-5}, 2014.

\bibitem{Kingma2017}
D.~P. Kingma and J.~Ba,
\newblock ``Adam: {A} {M}ethod for {S}tochastic {O}ptimization,''
\newblock in {\em Proc. of the 3rd Int. Conf. for Learning Representations, San
  Diego}, 2015.

\bibitem{klambauer2017selfnormalizing}
G.~Klambauer, T.~Unterthiner, A.~Mayr, and S.~Hochreiter,
\newblock ``Self-{N}ormalizing {N}eural {N}etworks,''
\newblock in {\em Proc. of the 30th Conf. on Neural Information Processing
  Systems (NIPS), Long Beach, CA, USA}, 2017, pp. 972--981.

\end{thebibliography}

\end{document}